\documentclass{aa} % for a referee version
\usepackage{graphicx}
\usepackage{caption}
\usepackage{amssymb}
\usepackage{enumerate}
\usepackage[toc,page]{appendix}
\usepackage{ulem}

\usepackage{booktabs}
\bibpunct{(}{)}{;}{a}{}{,}

\usepackage{natbib,twoopt}
\usepackage{hyperref} %% to avoid \begin{figure}[hbtp!]citeads line fills
\bibpunct{(}{)}{;}{a}{}{,}             %% natbib format for A&A and ApJ
\def\aj{AJ}%
\def\araa{ARA\&A}%
\def\apj{ApJ}%
\def\apjl{ApJ}%
\def\apjs{ApJS}%
\def\ao{Appl.~Opt.}%
\def\aap{A\&A}%
\def\aaps{A\&AS}%
\def\mnras{MNRAS}%
\def\pasp{PAGP}%
\def\nat{Nature}%
\def\gca{Geochim.~Cosmochim.~Acta}%
\begin{document}

   \title{CNO behaviour in planet-harbouring stars.
   \thanks{Based on observations collected with the UVES spectrograph at the 8-m Very Large Telescope (VLT) - program IDs: 074.C-0134(A), 075.D-0453(A), 086.D-0082(A), 093.D-0328(A), installed at the Cerro Paranal Observatory.}}

   \subtitle{I. Nitrogen abundances in stars with planets.}

   \author{L. Su\'arez-Andr\'es 
          \inst{1,2}\fnmsep\thanks{\textit{Send offprint requests to}: L. Su\'arez-Andr\'es
                \newline \email{lsuarez@iac.es}}
                        , G. Israelian    \inst{1,2},     
                        J.I. Gonz\'alez Hern\'andez \inst{1,2},
                        V. Zh. Adibekyan \inst{3},
                        E. Delgado Mena \inst{3},
                        N. C. Santos \inst{3,4}
                \and
                        S. G. Sousa \inst{3,4}
          }

   \institute{Instituto de Astrof\'isica de Canarias, E-38205 La Laguna,
Tenerife, Spain\\
              %\email{}
              \and
           Depto. Astrof\'isica, Universidad de La Laguna (ULL),
E-38206 La Laguna, Tenerife, Spain\\
        \and
          Instituto de Astrof\'isica e Ci\^encias do Espa\c{c}o, Universidade do Porto, CAUP, Rua das Estrelas, 4150-762 Porto, Portugal  \\
         \and Departamento de F\'isica e Astronomia, Faculdade de C\^iencias, Universidade do Porto, 4169-007 Porto, Portugal} 

   \date{Received X, 2016; accepted X, 2016}

% \abstract{}{}{}{}{} 
% 5 {} token are mandatory
 
  \abstract
   {Carbon, nitrogen, and oxygen (CNO) are key elements in stellar formation and evolution, and their abundances
should also have a significant impact on planetary formation and evolution.}
  {We present a detailed spectroscopic analysis of 74 solar-type stars, 42 of which are known to harbour 
  planets. We determine the nitrogen abundances of these stars and investigate a possible connection between N and the presence of planetary companions. }
  {We used VLT/UVES to obtain high-resolution near-UV spectra of our targets. Spectral synthesis of 
  the NH band at 3360\AA \space was performed with the spectral synthesis codes MOOG and FITTING.}
   {We identify several spectral windows from which accurate N abundance can be obtained. Nitrogen distributions for stars with and without planets show that planet hosts are nitrogen-rich when 
   compared to single stars. However, given the linear trend between [N/Fe] vs [Fe/H], this fact can be explained as being due to the metal-rich nature of planet hosts.}
   {We conclude that reliable N abundances can be derived for metal-rich solar type stars from the near UV molecular band at 3360 \AA. We confirm a linear trend between [N/Fe] and metallicity expected from standard models of Galactic chemical evolution. }


\titlerunning{Nitrogen abundances in planet harboring stars.}
\authorrunning{Su\'arez-Andr\'es, L. et al.}
%  \titlerunning{Nitrogen abundances in stars with planets.}
%
\maketitle
%__________\maketitle
%______________________________________________________
%
\section{Introduction}

The study of C, N, and O in stars is crucial because they are the most abundant elements after H and He. These 
elements play an important role in stellar interiors because they generate energy through the CNO cycle, thereby affecting the lifetime of the stars \citep{liang}

Nitrogen is created in a different nucleosynthetic process from that giving rise to carbon and oxygen. Whereas for carbon and oxygen the dominant 
production modes are the $\alpha$-chain reactions, for nitrogen the dominant production mode lies in the 
re-arrangement of nuclei during the CNO cycle \citep[see][]{maeder_book}. One important question regarding N 
is its origin. Since N needs C and O to be formed, if it is formed from pre-existing C and O in the star, it 
is called ``secondary''. If, instead, the carbon and oxygen are produced in the star itself and then used to produce 
nitrogen, then the nitrogen is called  ``primary''. 
Several studies have proved that production of nitrogen at low metallicities comes 
from primary rather than secondary sources  \citep{pagel,bessel82,carbon87,henry00, israelian04}. At higher metallicities, secondary processes dominate. 

Two main sources for primary production have been proposed:
\begin{enumerate}
        \item Rotating massive stars, implying detached N and Fe abundances and overabundance of N regarding Fe \citep{maeder00}. 
        \item Intermediate-mass stars (4-8 $M_{\odot}$) during their thermally 
pulsing AGB phase through CNO processing in the convective envelope
\citep{marigo01,van97}. 
The contribution by massive stars is negligible \citep{liang,pettini}, so dominant contributors are intermediate- 
and low-mass stars (ILMS). These ILMSs are also a source for secondary production.
\end{enumerate}

\citet{gonzalez97} and \citet{santos01} discovered that, on average, planet hosts are more 
metal-rich than ``single'' stars (stars with no known companion planet; from now on designated as single stars). 
Two hypotheses have been suggested to explain this anomaly:
\begin{enumerate}
        \item \textit{Self enrichment:} This scenario should be triggered by the action 
of hot jupiters migrating from outside the protoplanetary disc. During this migration, planetesimals from the 
disc are accreted on to the star. This mechanism would be efficient only during the first 20--30 Myr or later when 
the surface convection layers of the star for the first time attains its minimum size configuration \citep{chavero10}. 
\citet{israelian01, israelian03} found evidence of pollution, suggesting the infall of a planet, in HD82943\citep[see, 
however,][]{mandell04, ghezzi09}.
\newline
The self-enrichment scenario could lead to a relative 
overabundance of refractories, such as Si, Mg, Ca, Ti, and the iron-group elements, compared to 
volatiles, such as C, N, O, S, and Zn. The accretion of small planets may affect the composition
 of the convective layer of the stars \citep[see][and references therein]{theado}. To how great an extent this affects the accretion of volatiles such as C, N, and O \citep[see][]{jonay13} remains unanswered.
        \item \textit{Primordial cloud:} \citet{santos01, santos02, santos03} proposed 
that this over-abundance most probably is caused by a metal-rich primordial cloud. They claimed 
that the observed abundances are representative of the primordial cloud where the star was formed. This 
idea is supported by models of planet formation and evolution based on the core-accretion process \citep[e.g.][]{ida02, mordasini12}
\end{enumerate}
Later abundance studies on stars with and without planets confirmed the primordial cloud as the most likely reason for the metal-rich nature of planet-host 
stars \citep{santos04,santos05,valenti05}, although this correlation is valid for giant planet hosts only \citep{sousa08,sousa11b,buchhave}

Interestingly, recent studies suggest that specific element abundances may have a 
particularly relevant role in the planet formation process \citep{adi14, adi15} or in its 
composition \citep{santos15}. The abundances of volatiles (such as C, N, and O) may be particularly relevant in this respect. This consideration prompted the study of specific chemical abundances in the planet hosts.
 There are very few studies of nitrogen in solar-type stars owing to the lack of strong atomic lines.
 \citet{ecu04} studied the abundance of nitrogen of 91 solar type stars. Using both the NH molecular 
 band at 3360\AA\space and the NI atomic line at 7468\AA, they rejected the self-enrichment scenario 
 as a formation source on the grounds that they could find no underabundances of volatiles compared to refractory 
 elements. They showed that the [N/H] abundance  scales perfectly with metallicity for both planet-host and 
 comparison samples. They also found no difference for nitrogen in the [N/Fe] abundance ratios  when comparing stars with and without planets.
 
More recently, \cite{dasilva},  using the CN 
band at 4215\AA , have studied the abundance of nitrogen in 140 dwarf stars. They found a steeper slope for the [N/Fe] versus [Fe/H] abundance ratios than 
\cite{ecu04}.

The main problem in studying nitrogen abundances is the lack of strong nitrogen lines in the red 
part of the spectrum so that near-UV measurements are required. 
We are forced to study the very crowded molecular band at 3360\AA, where continuum determination is not straightforward.

The purpose of this study is to extend the sample of \citet{ecu04} and investigate in detail nitrogen 
abundances using the 3360\AA \space NH molecular band. We have performed a systematic study of  nitrogen 
abundances in dwarf stars with a wide range of stellar parameters. 
The strategy and methodology followed in this study is the same as that stated in \citet{ecu04}, but 
with a larger sample and higher-quality spectra.

We have also determined the kinematic properties of our sample. 
Then using the kinematic and chemical properties of the stars, we separated them into different stellar populations to investigate 
the N abundances within a Galactic context.

This work is the first step in comparing CNO abundances of planet-host stars with the CNO 
abundances of their planets (through transmission spectroscopy or direct spectroscopy, as in the case 
of \textit{Spitzer}). Testing and improving planetary formation models will play a key role in future studies of habitability, CNO being key elements for life. 

\section{Sample description}

The high-resolution spectra analysed in this study were obtained with the UVES spectrograph installed at 
the VLT/UT2 Kueyen telescope (Paranal Observatory, ESO, Chile) during several campaigns (see Table \ref{logs}) 
and have previously been  used in the analysis of stellar parameters, together with the derivation of precise 
chemical abundances \citep[see e.g.][]{santos04, sousa11a, sousa11b, elisa}.

\begin{table}[!h]
        \caption{Observing logs for the campaigns}
        \label{logs}

        \centering
        \begin{tabular}[l c ]{ l  c  }
                \hline
                \hline
                Campaign & Observing dates \\

                \hline 
                074.C-0134(A)     &    21-22 Dec. 2004          \\
                075.D-0453(A)     &    First semester 2005      \\  
                086.D-0082(A)    &    Oct. 2010 - Nov. 2011   \\
                093.D-0328(A) &   Mar. - Jul. 2014 \\
                \hline \\
        \end{tabular}
        
        %       \end{center}
        
\end{table}
High spectral resolving power ($R=80,000$) and a relatively high signal-to-noise ratio (S/N) are optimal for properly analysing the NH band at 3360\AA. The average S/N of our sample in the studied regions is 150.

The sample consists of 74 dwarf stars (42 stars with planets\footnote{Data from: \textit{www.exoplanet.eu}.} and 32 
comparison stars without detected planets) with effective temperatures between 4583 K and 6431 K, metallicities from 
$-0.45$ to $0.55$ dex, and surface gravities from $3.69$ to $4.82$ dex. Comparison sample stars were taken from HARPS and CORALIE samples. 

\section{Analysis}

\subsection{Stellar parameters and chemical abundances}

The stellar parameters used in this study were taken from \citet{sousa08,sousa11a, sousa11b, tsantaki} and \citet{jonay10, jonay13}.
All stellar parameters used were derived by measuring equivalent widths of Fe \rm{I} and Fe \rm{II} lines using the code ARES \citep{ares}. Also, chemical abundances of elements other than N were adapted in targets with more recent stellar parameters (derived by our group with the same technique). To do this, the uncertainties presented in their original source were followed so systematic effects in the chemical abundances or the stellar parameters are negligible, not affecting the consistency of our results.

 Chemical abundances of the elements with spectral lines present in the NH band were obtained from \citet{adi12} 
 and \citet{jonay10, jonay13}. These elements are Ca, Ti, Mn, and Si (note that the abundances of these elements were simply scaled with iron in \cite{ecu04}). 
The N abundance is also affected by C and O molecular equilibrium. The C and O abundances were obtained from 
Su\'arez-Andr\'es et al. (2016, in prep) and \citet{sara}, respectively. However, there are 26 stars with no previous measurements of O. We have decided to use the aforementioned
results to interpolate for a given set of  stellar parameters and use these new O abundances to 
calculate the molecular equilibrium. Our tests show that N abundances are unaffected by changes of the order of $\pm$0.2 dex in C and O.

Nitrogen abundances were determined using a standard local thermodynamic equilibrium (LTE) analysis with the spectral 
synthesis code MOOG \citep[2013 version]{sneden} and a grid of Kurucz (1993) ATLAS9 atmospheres. All the atmospheric parameters, 
$\rm T_{\rm eff}$, $log g$, $\xi_{t}$ and [Fe/H] were taken, as already mentioned, from \citet{sousa08, sousa11a, sousa11b, tsantaki} 
and \citet{jonay10, jonay13}. The adopted solar abundances for nitrogen and iron were 
log$\epsilon(\rm N)_{\odot}$=8.05 dex and log$\epsilon(\rm Fe)_{\odot}$ = 7.47 \citep{santos04}.

\begin{figure}[!hbt]
        
        \scalebox{1.2}[1.]{
                
                \includegraphics[angle=90, width=0.8\linewidth]{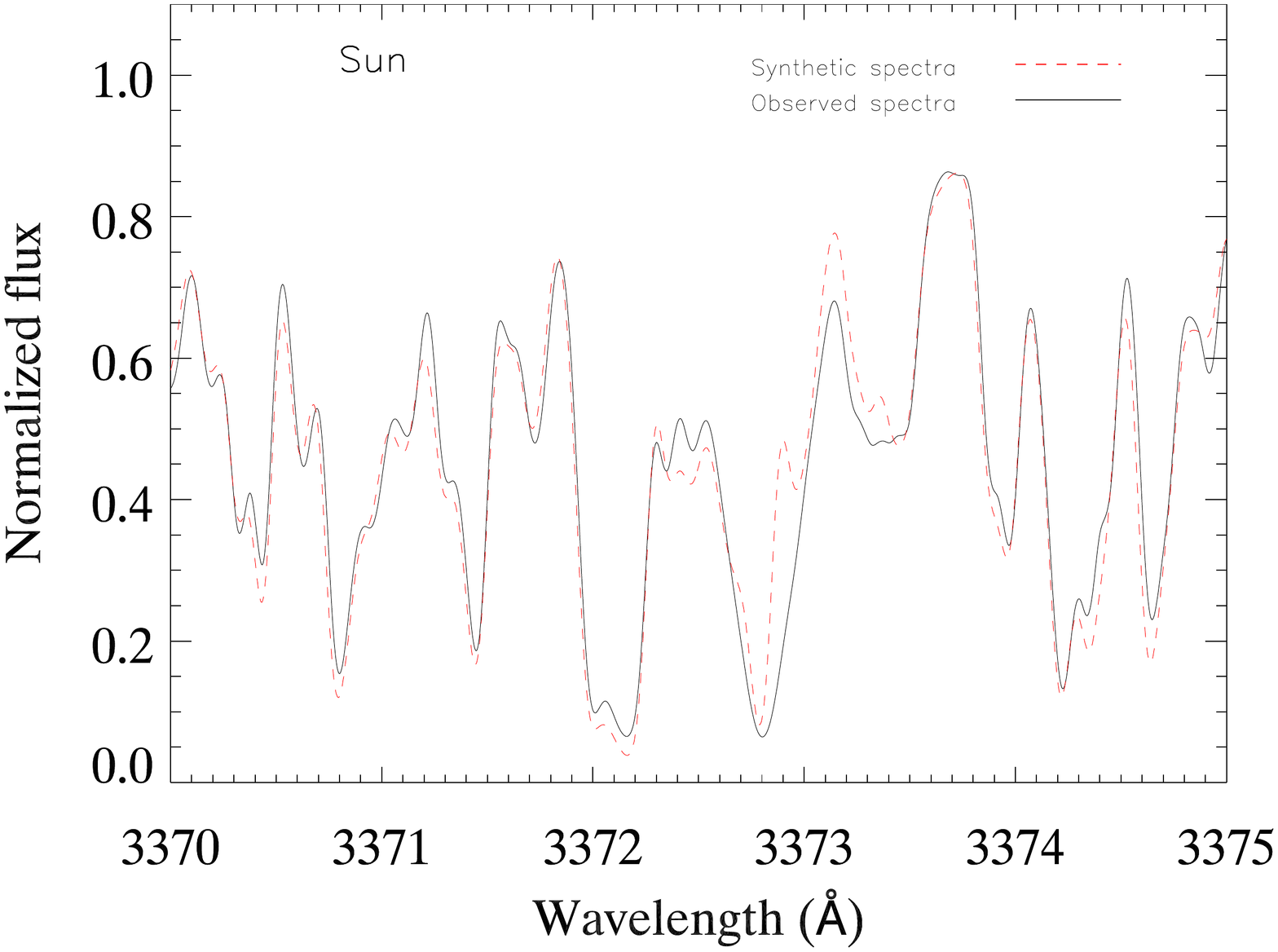}}

        \scalebox{1.2}[1.]{
                \includegraphics[angle=90,width=0.8\linewidth]{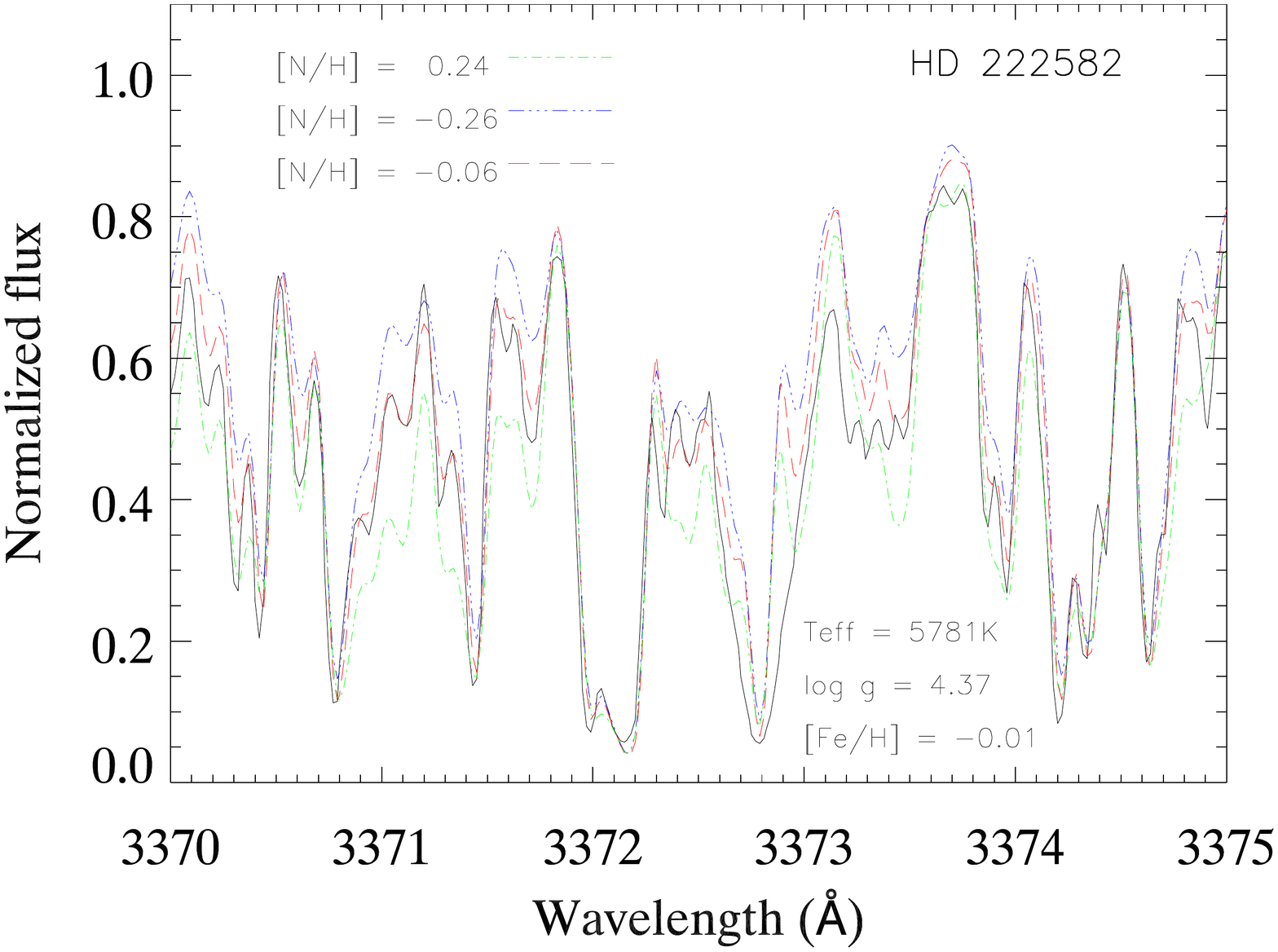}}
        
        \scalebox{1.2}[1.]{
                \includegraphics[angle=90,width=0.8\linewidth]{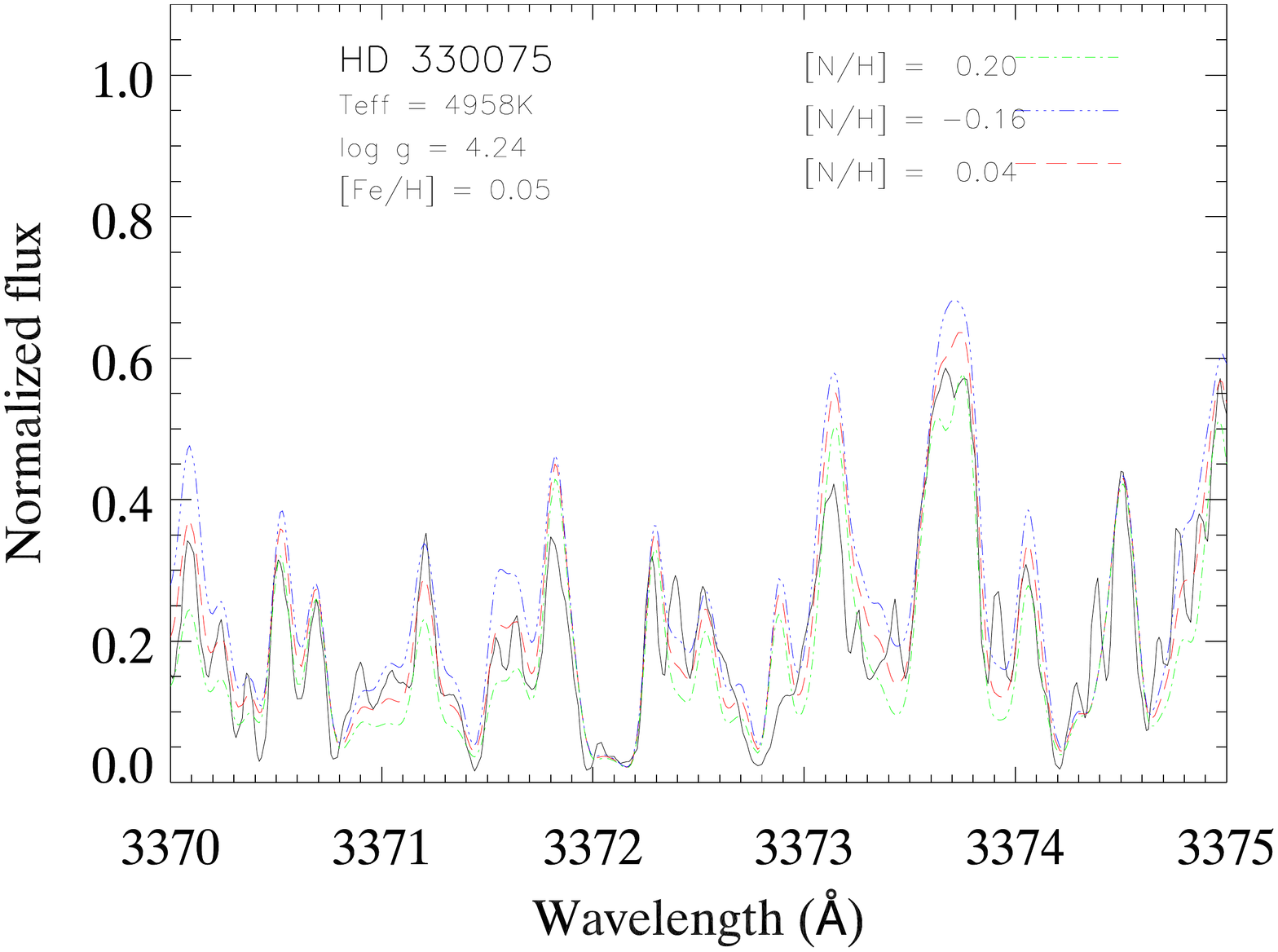}}
        
        \caption{\textit{Top panel:} Solar observed (solid line) and synthetic (dotted line) spectra in the 
                spectral region $\lambda \lambda$3370-3375\AA. \textit{Middle and bottom panel:} Observed (solid) and 
                synthetic spectra (green-dotted lined, red-dashed and blue-dashed dotted) of HD 222582 and HD 330075.}
        \label{fitting}
        
\end{figure}
\subsection{The NH band}

The NH band is the strongest feature observed in the $\lambda \lambda3345-3375 \AA$ spectral region. We determined N abundances by fitting synthetic spectra 
to the data in this wavelength range. 
The dissociation potential used for NH spectra is $D_{o}=3.37$ eV, as recommended in \citet{grevesse90}. 
The complete line list used in this study was obtained from \citet{ecu04}, which was calibrated with the 
KURUCZ ATLAS spectrum \citep{kurucz} using the abundance  log$\epsilon(\rm N)_{\odot}$=8.05 dex.

The number and the strength of the atomic lines of different elements in the spectral 
region of the NH band increase with metallicity. The presence of many blended lines and 
 numerous strong molecular bands make the placement of the continuum level 
very difficult. The  high-resolution solar atlas \citep{kurucz} can help to achieve 
a reliable continuum placement for solar-type stars (e.g.\ \cite{ecu04}). 
However, large  variations of metallicity and effective temperature among the stars 
in our sample do not allow us to use the solar spectrum as a reference. 
To account for this effect, we have generated a grid of synthetic spectra 
for all the stars in our sample. Synthetic spectra are calculated for a 
given set of stellar parameters and variations of nitrogen abundances
[N/H] between $-0.7$ and 0.5. Our reference points for the continuum placement 
are those for which the flux variation (for a given $T_{\rm eff}$) due to the N abundance 
changes from [N/Fe] = $-0.7$ to +0.5,  is less than 1\%.  This strategy is 
demonstrated in Fig. \ref{fitting} for two stars with different atmospheric parameters.
Rotational broadening was set as a free parameter (it was fixed in \cite{ecu04}) 
with $v\sin i$ varying between 0.0 and 14.0 with a step of 1 km/s. Macroturbulence was not
taken into account. In order to find the best fit abundance value for each star, we used 
the FITTING program  \citep{jonay11} and the MOOG synthesis code  in 
its 2013 version. The best fit was obtained using a $\chi^{2}$ minimization procedure by 
comparing each synthetic spectrum with the observed one in the following spectral regions:  $\lambda\lambda3344.0-3344.3\AA$,
$\lambda\lambda3346.2-3346.7\AA$,
$\lambda\lambda3347.0-3347.8\AA$,
 $\lambda\lambda3353.8-3354.4\AA$, $\lambda\lambda3357.4-3358.0\AA$, $\lambda\lambda3358.5-3359.7\AA$, 
 $\lambda\lambda3360.3-3362.0\AA$, $\lambda\lambda3364.1-3364.8\AA$, 
 $\lambda\lambda3370.8-3371.8\AA$,
  $\lambda\lambda3374.9-3375.4\AA$.
These spectral regions were chosen because of the presence of relatively strong NH features that allow reliable abundance measurements.
We use a $\chi^{2}$ comparison for the observed and synthetic spectra, and we define $\chi^{2}$=$\sum_{i=1}^{N}(F_{i}-S_{i})^{2} / N $, where $F_{i}$ and $S_{i}$ 
are the observed and synthetic fluxes respectively at wavelength point $i$.  
The  $\chi^{2}$ for each spectral region was normalised with the number of points, $N$. 
Best-fit nitrogen abundances are extracted from each spectral range and  the final nitrogen abundance 
for each star is then computed as the average of these values.

The FITTING program  creates the following input data, required by MOOG: the atmospheric model, the line list, 
the range of nitrogen abundance of the grid of synthetic spectra, and the wavelength at which we want to analyse 
these abundances. To obtain the final abundance value, we analysed ten different ranges but  use  only
those that have abundance values within 1$\sigma$. 

In Fig.~\ref{fitting} we show the observed and synthetic spectra for the Sun and two stars that are
 depicted for different temperatures and metallicities within our sample. For these two stars, three different nitrogen abundances are also shown.

To examine how variations in the atmospheric parameters affect the NH abundances, we test [N/H] sensitivity in stars with very 
different parameter values, given the wide range of stellar parameters. For each set of stars we tested the nitrogen-abundance sensitivity to changes in the atmospheric 
parameter ($\pm$ 100K for $\rm T_{\rm eff}$, $\pm$ 0.2 dex in $log g$, $\pm$ 0.2 in metallicity). The results are shown in 
Table \ref{sens}. The effect of microturbulence was not taken into account because an increase of 0.3 $km s^{-1}$ produced 
an average decrease of 0.002 dex in nitrogen abundance, which is negligible in comparison with the effects of other 
parameters. The error due to continuum placement of 0.1 dex was considered for all stars. All effects were added quadratically 
to obtain the final uncertainties in nitrogen abundances using the following relation:

$\Delta {\rm [N/H]} = (\Delta_\sigma^2+\Delta_{T_{\rm eff}}^2+\Delta_{\log g}^2+\Delta_{\rm met}^2+\Delta_{\rm cont}^2)^{1/2}$

\section{Results}

We analysed near-UV high-resolution spectra of 42 planet host stars and 32 comparison stars, as 
mentioned in Section 2. We aim to explore possible differences in nitrogen abundances between 
the two samples. Our results for planet-host are presented stars in Table~\ref{tablacon} and for the comparison sample in Table~\ref{tablasin}.

\begin{figure}[!t]
        \begin{center}
                \scalebox{1.2}[1.15]{
                        \includegraphics[angle=0,width=0.8\linewidth]{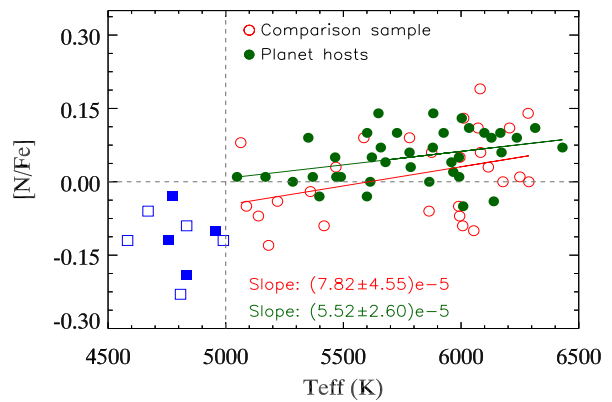}}
        \end{center}
        \caption{[N/Fe] abundance ratio of the stars in this study plotted against effective temperature, $\rm T_{\rm eff}$. 
        Filled green circles represent planet hosts and  open red circles the comparison sample. Filled blue squares represent cool planet hosts while open blue squares represent single stars. The vertical dashed 
        line at 5000 K separates the cool stars from the studied sample.}
        \label{teff}
\end{figure}

In Fig.~\ref{teff} we show the [N/Fe] abundance ratio as a function of $\rm T_{\rm eff}$. Stars with effective 
temperatures below 5000 K were excluded from the analysis because of uncertainties in the behaviour of those cool 
stars: we find no explanation for the decrease in nitrogen abundance as we move to lower temperatures. The vertical dashed line at 5000 K separates the 
cool stars from the sample studied. From now on in this paper, all results and conclusions will refer to stars with 
$\rm T_{\rm eff}$ > 5000 K (65 stars). Because of the high dispersion, we did not find any clear trend of [N/Fe] with $\rm T_{\rm eff}$. 
However, it can be seen that most of the planet-host stars have [N/Fe] > 0.

 \begin{figure}[!h]
        \begin{center}

                \centering
                \scalebox{1.3}[1.2]{
                        \includegraphics[angle=180,width=0.8\linewidth]{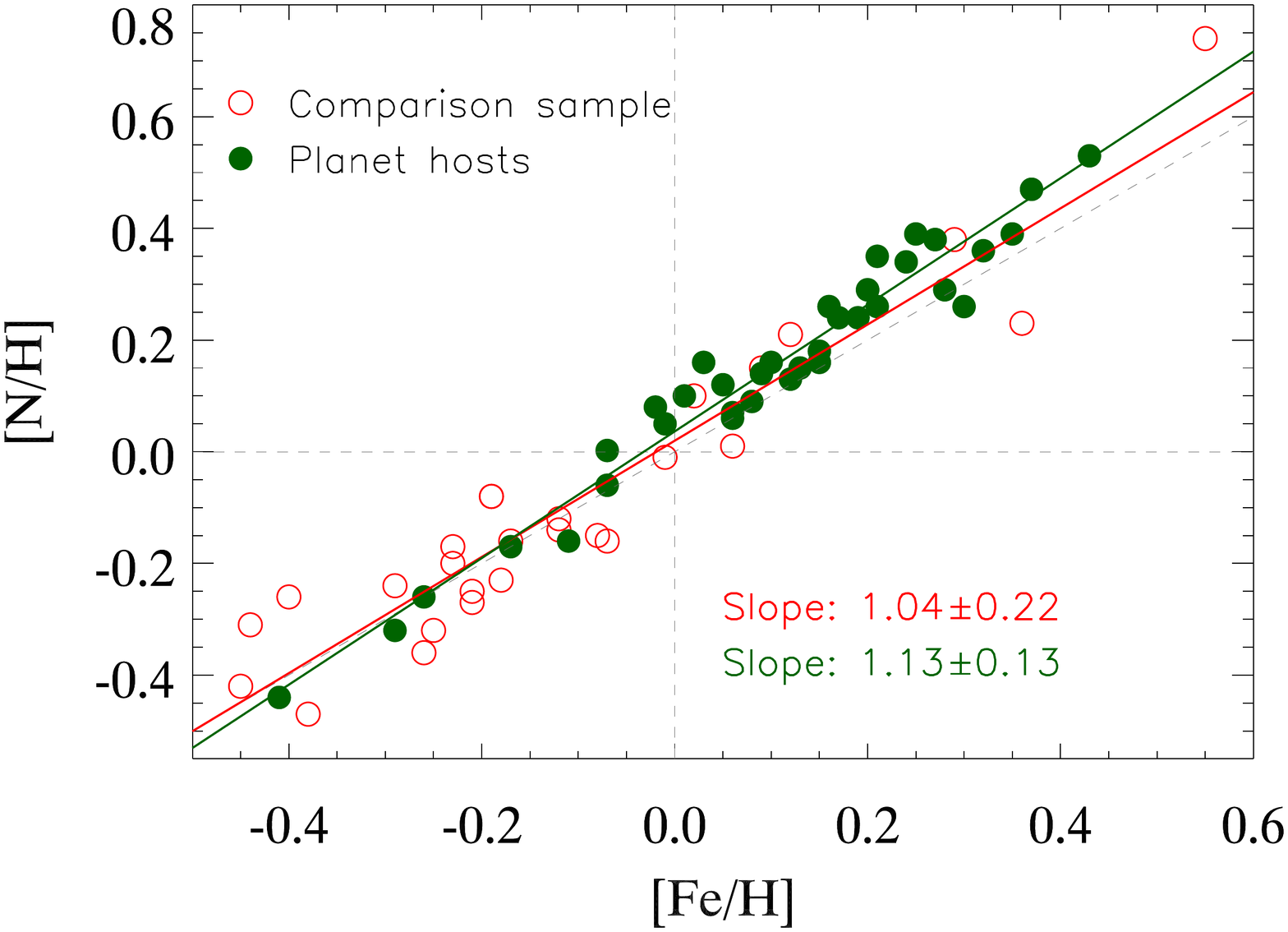}}
                
                \scalebox{1.23}[1.15]{
                        \includegraphics[angle=0,width=0.85\linewidth]{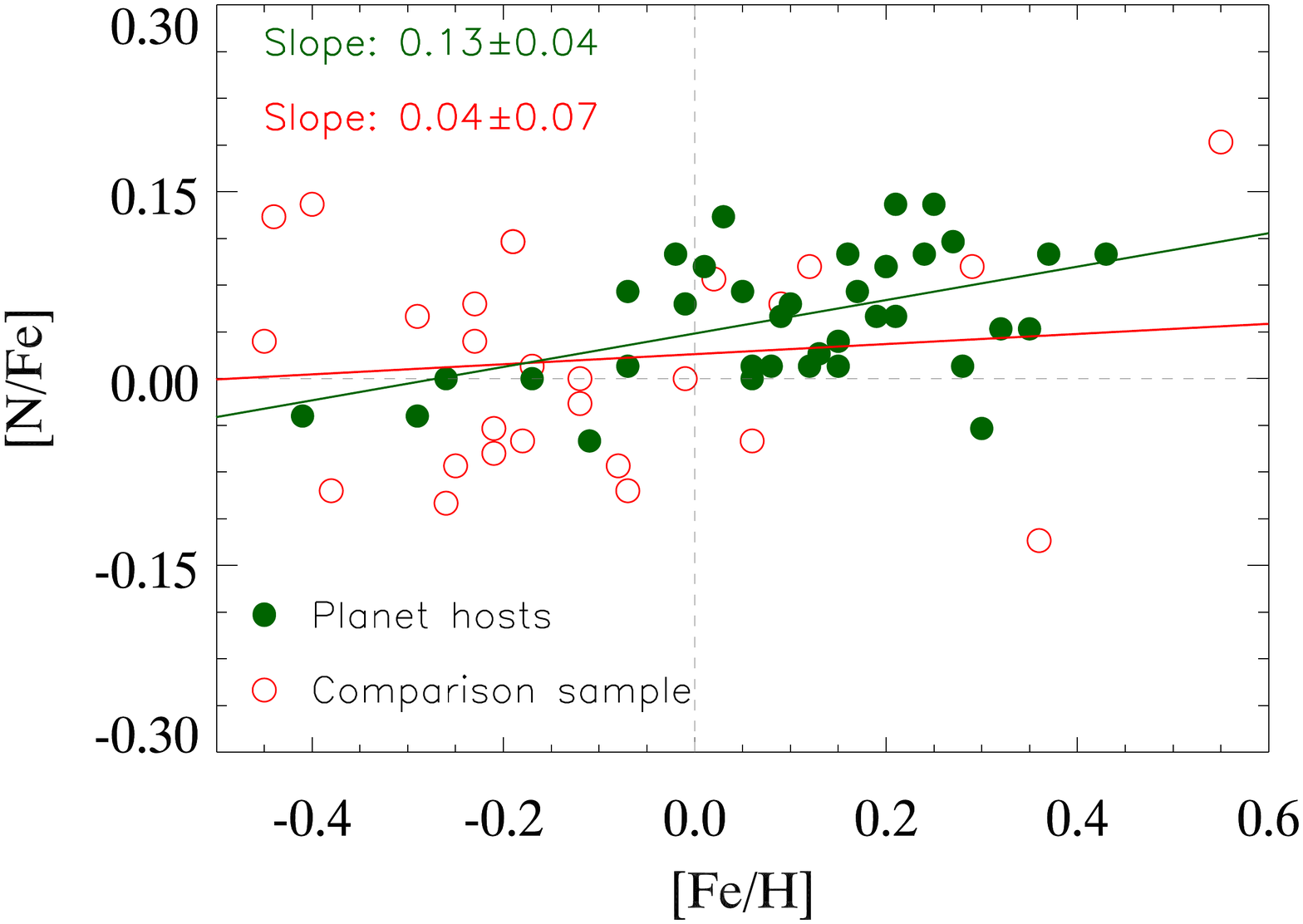}}
                \caption{[N/H] and [N/Fe] abundance ratios of our sample stars versus metallicity, [Fe/H]. Filled 
                        green circles represent planet hosts and open red circles, the comparison sample. Dashed lines stand for solar values.}
                \label{met1}
                
        \end{center}
        
 \end{figure}

We have also looked for distinguishable trends between these samples by representing [N/Fe] and [N/H] abundance ratios as 
functions of [Fe/H] for both samples (see Fig.~\ref{met1}). These plots indicate that both samples behave approximately 
similarly. However, there seems to be a steeper trend in [N/H] vs [Fe/H] for stars with planets, whereas stars 
without planets almost maintain the 1:1 relation. We note the same behaviour in the bottom panel of Fig.\ \ref{met1} ([N/Fe] 
vs [Fe/H]), where the stars with planets below and above solar metallicity have values of [N/Fe] lower and higher 
than zero respectively. Unfortunately, owing to the metal-rich nature of giant-planet hosts, the number of stars with 
giant planets at metallicities below solar is too small for us to be able to confirm this behaviour. The observed 
trend may simply be related to  Galactic chemical evolution.

\begin{figure}[!hbtp]

        \begin{center}
        \scalebox{1.2}[1.1]{
   \includegraphics[angle=0,width=0.8\linewidth]{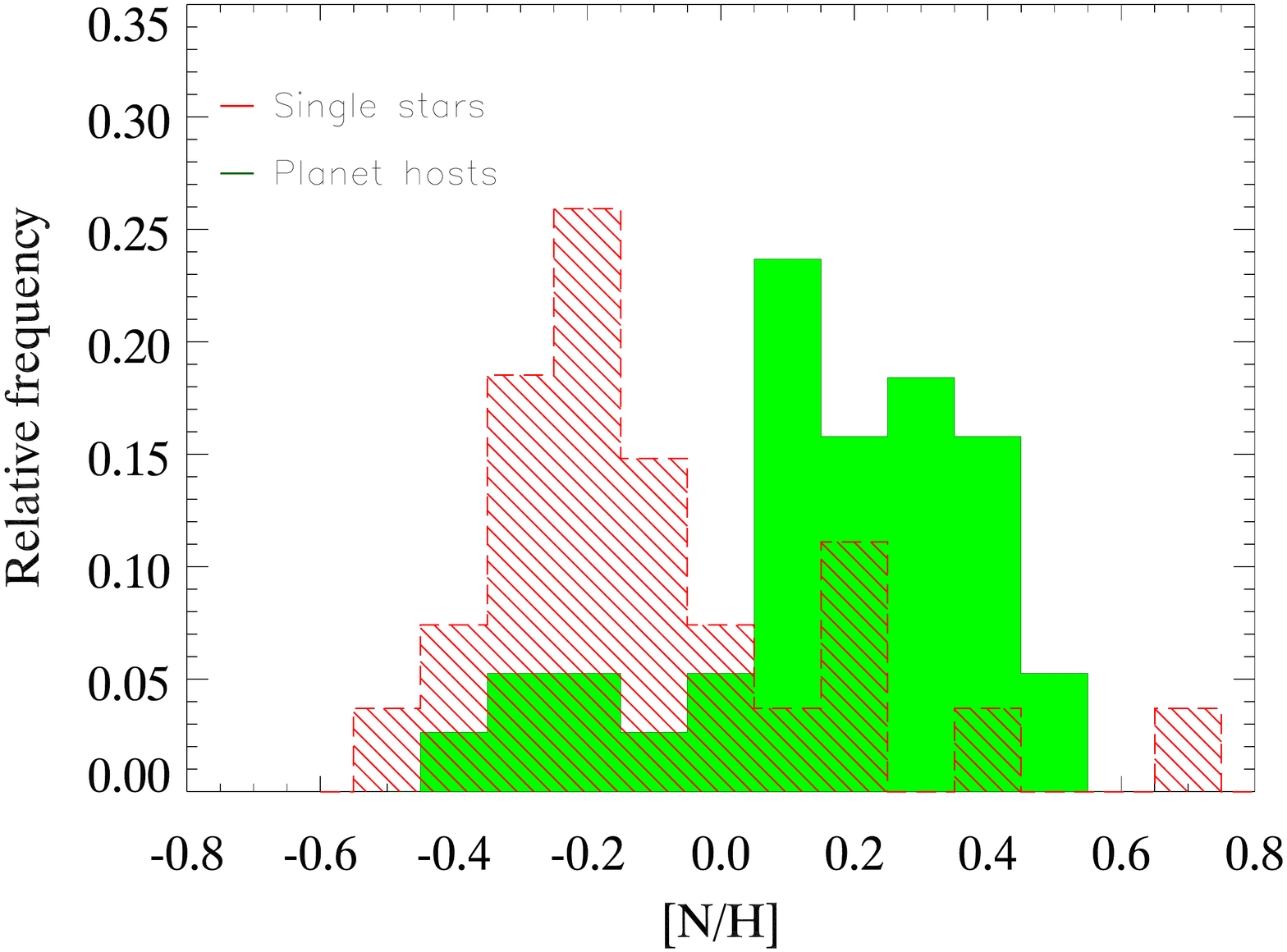}}
        \scalebox{1.2}[1.1]{
   \includegraphics[angle=0,width=0.8\linewidth]{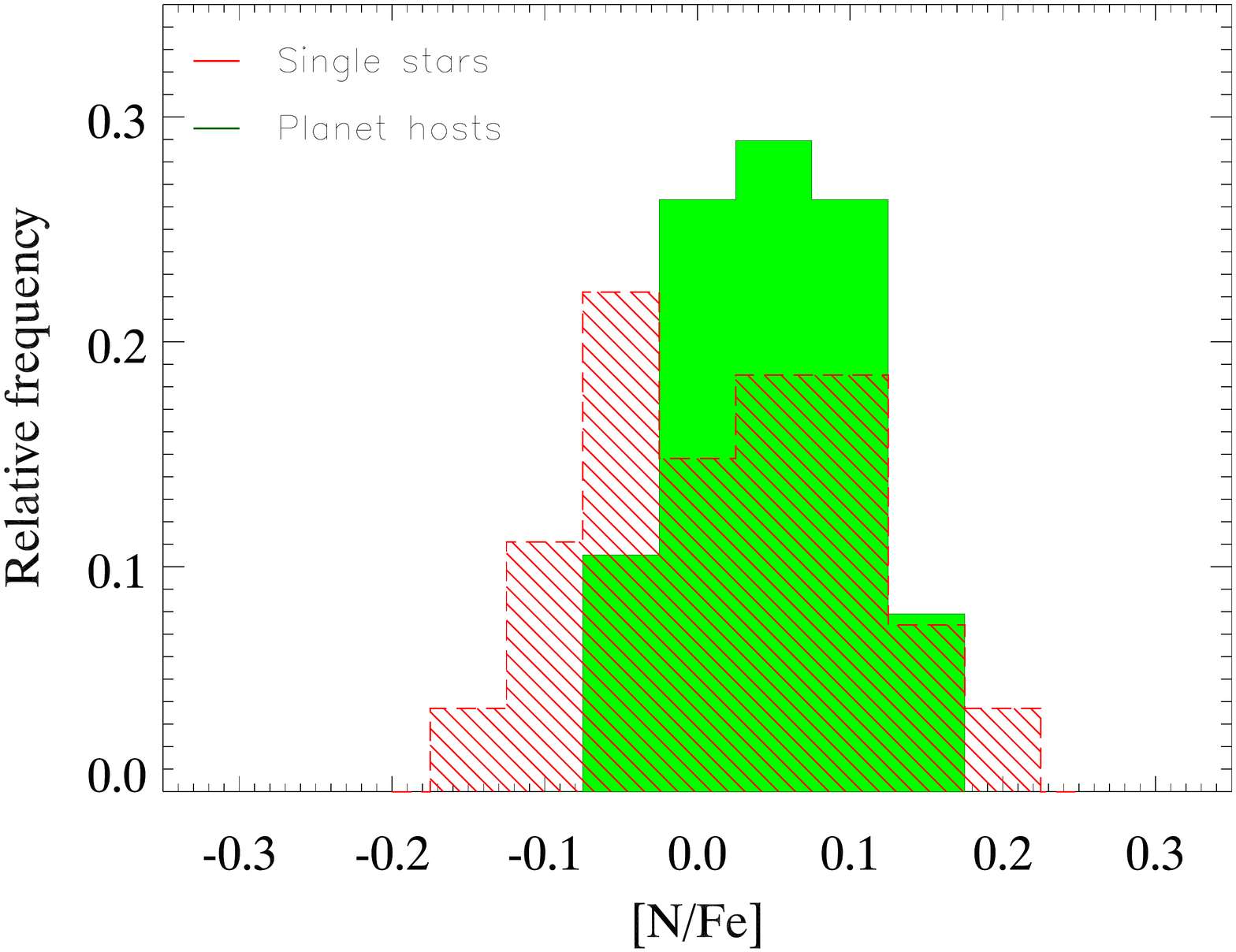}}
        \end{center}
        \caption{[N/H] and [N/Fe] abundance distributions.}
        \label{histo}
\end{figure}

In the top panel of Fig.~\ref{histo} we show the [N/H] abundance distributions for both samples. As can be seen, there is an 
offset between the samples, which is  expected because planet-host stars are metal rich as compared with single stars. We expect this result because, if nitrogen scales with iron, then we can expect higher [N/H] because giant-planet host stars are enhanced in Fe.
In the bottom panel of Fig.~\ref{histo} we show the [N/Fe] abundance distribution. We see that most of the stars with planets have [N/Fe] $\geq$0 ($\sim 90$ per cent), as opposed to the single-star sample, more spread than the stars with planets sample. In this case, only $\sim$60 per cent of the stars have [N/Fe] $\geq$ 0. 
A Kolmogorov-Smirnov (K-S) test predicts the $\sim$0.06 probability ($P_{KS}$) that stars with and without planets come from the same distribution. The distribution of abundance ratios  [N/Fe] vs [Fe/H] of stars without planets shows a weak linear increasing trend (see Fig. \ref{met1}), although the slope is consistent with zero. The number of points may not be sufficient to really confirm this increasing trend. Thus, we may conclude that stars with planets are, on average, nitrogen rich when compared to single stars, but it may be due to the metal-rich nature of the planet host stars.

\begin{figure}[!htbp]
        \scalebox{1.3}[1.2]{
                \centering
                {       \includegraphics[angle=180, width=0.8\linewidth]{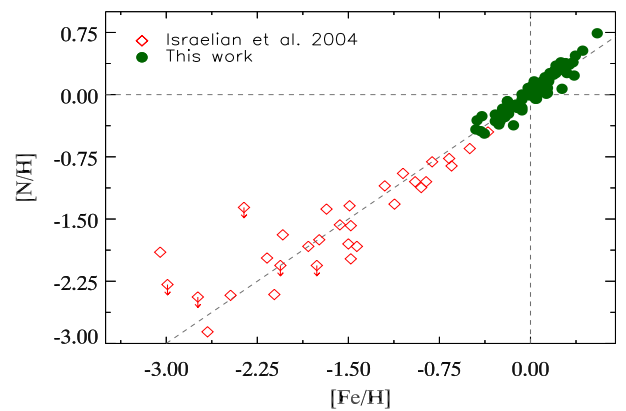}}}
        \caption{N/H plotted against Fe/H for metal-poor stars from \citet{israelian04} indicated by red squares and stars from this study indicated by green circles.}
        \label{metalpoor}
\end{figure}

If we extend the study of nitrogen abundance to metal-poor stars \citep{israelian04}, we will ensure 
that the behaviour seen in Figure \ref{met1} is part of the same trend observed in these metal-poor stars down to 
metallicites $\sim$ $-2$ $dex$. In Figure \ref{metalpoor} we can see how these two sets, (\cite{israelian04} and this study) 
follow the same trend down to metallicities of $-2.0$ $dex$ because nitrogen  behaves like a secondary production element in this range of metallicities. At metallicities lower than $-2.0$ $dex$, we see a signature of primary N.

\begin{table*}[!t]
        \caption{Sensitivity of the nitrogen abundance derived from the NH band at 3360\AA. Changes of 100 K in 
                $\rm T_{\rm eff}$, 0.2 dex in gravity, and 0.2 in [Fe/H] were applied.}             
        \label{sens}      
        \centering   
        \scalebox{1}[1]{      
                \begin{tabular}{ c c c c } 
                        \hline
                        \hline
                        & & \multicolumn{1}{c}{Star}\\
                        & & \multicolumn{1}{c}{($\rm T_{\rm eff}$; $\log g$; [Fe/H])}\\
                        \cmidrule{2-4} 
                        
                        &     & & \\
                        & HD 93083 & HD 222582 & HD 39091\\
                        & (5048; 4.32; 0.04) & (5779; 4.32; -0.01)& (6003; 4.42; 0.09) \\

                        \cmidrule{2-4}        
                        $ \Delta T_{\rm eff} = \pm$ 100K  & $\pm0.05$ & $\pm0.08$& $\pm0.10$ \\  
                        
                        \hline
                        &     & & \\
                        & HD 11964A & HD 93083 & HD 1237\\
                        & (5332; 3.90; 0.10) & (5105; 4.43; 0.09)& (5514; 4.50; 0.07)  \\ 
                        \cmidrule{2-4}                    
                        
                        $ \Delta \rm log g$ = $\pm$ 0.2 dex    & $\mp0.01$ & $\mp0.05$& $\mp0.03$\\
                        
                        \hline
                        &     & & \\
                        
                        & HD 4208  & HD 69830& HD 73256\\
                        & (5599; 4.44; -0.28) & (5402; 4.40; -0.06)& (5526; 4.42; 0.23)  \\                    
                        \cmidrule{2-4}   
                        
                        $ \Delta \rm([Fe/H]) =\pm$ 0.2 dex  & $\mp0.16$     & $\mp0.01$&$\mp0.08$\\
                        
                        \hline                
                \end{tabular}}
        \end{table*}

\section{Galactic evolution of N: dependance on age}

Interpretation of chemical abundances in terms of stellar ages can be helpful to constrain the effects of Galactic chemical evolution. 

Stellar ages for our sample were estimated applying stellar evolutional models from the Padova group \citep{padova}, using the web interface \footnote{http://stev.oapd.inaf.it/cgi-bin/cmd}\citep[see][for more details]{sousa11a}. 

\citealp{nissen} performed a detailed study of abundance ratios of several elements, such as C, O and Si, as a function of the stellar age. They conclude that each element behaves in a different way regarding the stellar age, suggesting that more variables such an evolving initial mass function and asymptotic giant branch stars should also be considered.
Although that work is only for solar twin stars, we try to extend this result to solar type stars and for nitrogen, an element not studied in that work.

\begin{figure}[!htbp]
        \scalebox{1.27}[1.15]{
                \centering
                {       \includegraphics[angle=180, width=0.8\linewidth]{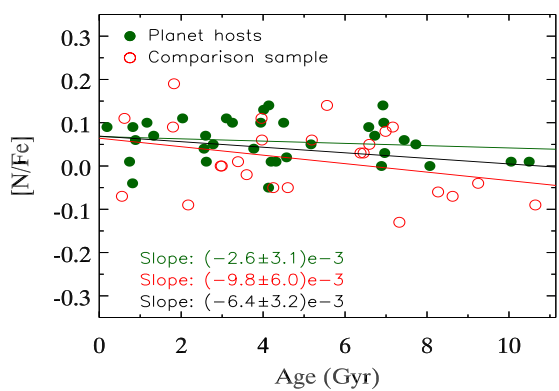}}}
                \caption{Abundance ratio of [N/Fe] as a function of stellar age for the stars belonging to the thin disk. Linear fit is provided for the whole sample (black line) and each sub-sample (red and green lines).}
        \label{age}
\end{figure}

In the top panel of Fig.\ref{age} we show [N/Fe] abundances as a function of stellar age for the thin disk stars. There seem to be a weak trend in the behaviour of nitrogen, as its abundance decreases with age. Although single stars appear to show a steeper decreasing trend towards older stars than planet hosts stars, probably due to the low number of points and the high dispersion of the abundance measurements,the slopes of these two trends are consistent within their error bars.

\section{Kinematic properties}

To study the kinematic properties of the sample stars and which stellar populations they belong to, we applied both purely chemical 
\citep[e.g.][]{adi11, recio14} and kinematic approaches \citep[e.g.][]{bensby03, reddy06}. The Galactic space velocity components of the stars
were calculated as in \citep[][]{adi12} using the astrometric \footnote{The SIMBAD Astronomical Database (http://simbad.u-strasbg.fr/simbad/) was used.} 
and radial velocity data of the stars. The average errors in the $U$, $V$, and $W$ velocities are approximately 2--3 km s$^{-1}$.

To assess the likelihood of the stars being members of different stellar populations we followed \citet{reddy06}, and adopted the results of
\citet{bensby03} 
for the population fractions. According to this separation, among the 65 stars, we have 58 (89 per cent) stars from the thin disc, four from the thick disc, and three
transition stars that do not belong to any group.

\begin{figure}[!htbp]
        \scalebox{1.45}[1.5]{
                \centering
                                {       \includegraphics[angle=180,width=0.7\linewidth]{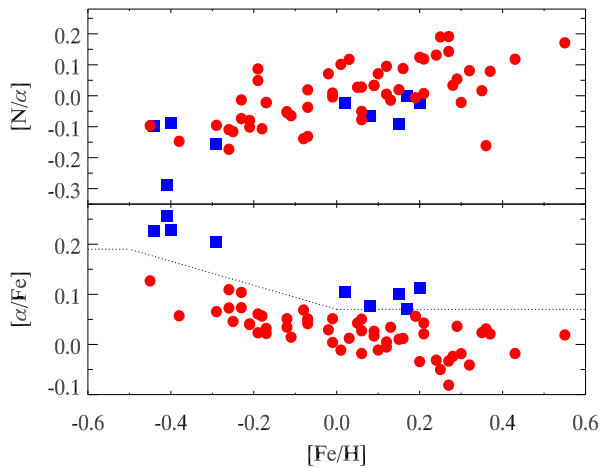}}}
        \caption{[N/$\alpha$] plotted against [Fe/H] (top) and [$\alpha$/Fe] plotted against [Fe/H] (bottom) for the sample. Stars that are enhanced in $\alpha$-elements are shown
        in blue squares and the thin-disc stars (non-$\alpha$-enhanced) are represented by red filled circles.}
        \label{N_Alpha_Fe}
\end{figure}

The separation of the Galactic stellar components based only on stellar abundances is probably superior to that based on kinematics alone \citep[e.g.][]{navarro11,
adi11} because chemistry is a relatively more stable property of 
sunlike stars than spatial positions and kinematics. 
We used the 
position of the stars in the [$\alpha$/Fe]--[Fe/H] plane (here $\alpha$ refers to the average abundance of Si and Ti) to separate the thin- and thick-disc stellar components. We adopt the boundary (separation line) between the stellar populations from \citet{adi11}. According to our separation, 53 stars 
($\approx$82 per cent) are not enhanced in $\alpha$-elements and belong to the thin-disc population. The [$\alpha$/Fe] versus [Fe/H] plot 
for the sample stars is shown in the bottom panel of Figure \ref{N_Alpha_Fe}.

In the top plot of Fig. \ref{N_Alpha_Fe} we show the dependence of [N/$\alpha$] on the metallicity. 
The figure shows that the [N/$\alpha$] abundance ratio correlates with the metallicity. This trend is expected if the
N abundance scales with the iron abundance, as was suggested above.

\section{The star--planet connection}

The increasing number of exoplanets discovered via different methods has led to a wide diversity of 
parameters (masses, radii, eccentricity, orbital period, etc.) being known for these planets.
Many studies have suggested that the formation of giant planets correlates with the metallicities of stellar 
hosts, \citep[see][]{santos01, santos04, fischer, sousa11a,mortier}. Moreover, it has been shown that the 
formation of planets (of both high and low masses) at low metallicities is favoured by enhanced $\alpha$-elements \citep{adi_over, adi12c}. 

\begin{figure}
        \centering
        \scalebox{1.4}[1.35]{
                \includegraphics[angle=0,width=0.7\linewidth]{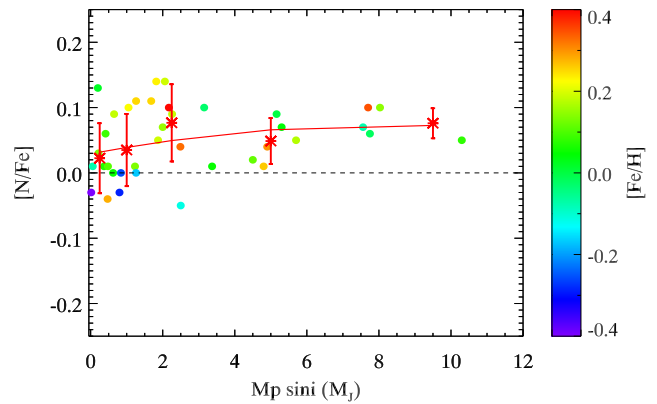}}
        \caption{[N/Fe] plotted against $M_{p}\sin i$, with [Fe/H] in the auxiliary axis. Dots represent the whole sample 
        and red asterisks represent binned values. We also represent in red a second-degree polinomial fit for those binned values.}
        \label{planet_mass}
\end{figure}

Models of planet formation require precise stellar abundances. Chemical abundances of planet hosts are useful 
when studying the properties of the planetary companion \citep{sousa15}. The relationship between [N/Fe] and the mass 
of the planet is shown in Figure \ref{planet_mass}. In those stars which host many planets, the most massive planet 
was considered in the plot. As we can see, the masses of the planets are between 0.015 and 10.3 $M_{J}$, but most 
stars have planets with less than 4 $M_{J}$ orbiting around them. To remove this bias, which is due to the lack of planets with 
higher masses, we created bins with increasing steps, with sizes of 0.5, 1, 1.5 $M_{J}$ and two bins of 4 $M_{J}$. 
Error bars indicate the standard deviation of each bin. Although the results are biased because of the lack of data in the highest-mass part, 
we cannot obtain a clear relationship between [N/Fe] and the planetary mass (covariance $s_{x,y}=0.031$; where $x=M_{J}$ and $y=[N/Fe]$).
We see an increase of the [N/Fe] ratio in the first mass bins followed by constant [N/Fe] value for the more massive planets. Even so, the number 
of stars is not statistically significant to confirm this. 

\section{Discussion and conclusions}

We present nitrogen abundances for 74 solar-type stars observed with the UVES spectrograph at the VLT/UT2 Kueyen 
telescope (Paranal Observatory, ESO, Chile). In our sample, 42  of the 74 are planet-hosts and 32 stars do not
 have any known planetary companion. All targets have been studied using spectral synthesis in the NH band at 3360\AA.
  The stars in our sample have effective temperatures between 4583 K and 6431 K, metallicities from $-0.45$ to $0.55$ dex, 
  and surface gravities from $3.69$ to $4.82$ dex.
We performed a detailed analysis of this sample to obtain precise nitrogen abundances and investigate possible differences 
between the stars with planets and the single stars. Nitrogen abundances of both samples behave in a similar way regarding $\rm T_{\rm eff}$, $\log g$, and [Fe/H].

 We extended our results to the metal-poor regime, comparing our results with previous work by \citet{israelian04}. Both 
 results can be accommodated under a common fit in an [N/H] versus [Fe/H] plot until metallicities down to $-2.0$ $dex$ (where the production of nitrogen has a secondary origin), suggesting that both metal-poor stars and solar-like stars follow the same behaviour.

The correlation between the presence of planets and nitrogen 
abundances is expected: The large amount of nitrogen in a protoplanetary disc would favour the formation of massive giant planets through the core accretion scenario. 

We searched for correlations between planet mass and [N/Fe] abundance ratio, but did not find any significant correlation.

\begin{acknowledgements}
        J.I.G.H. acknowledges financial support from the Spanish Ministry of Economy and Competitiveness (MINECO) under the 
        2011 Severo Ochoa program MINECO SEV-2011-0187 and the 2013 Ram\'on y Cajal programme MINECO RYC-2013-14875, and the 
        Spanish ministry project MINECO AYA2014-56359-P. E.D.M and V.Zh.A. acknowledge the support from the Funda\c{c}\~ao para 
        a Ci\^encia e Tecnologia, FCT (Portugal) in the form of the grants SFRH/BPD/76606/2011 and SFRH/BPD/70574/2010 
        respectively. V.Zh.A also aknowledges the support from COST Action TD1308 through STSM grant with reference Number: COST-STSM-TD 1308-32051. N.C.S. and S.G.S.also acknowledge support from FCT through Investigador FCT contracts of reference IF/00169/2012 and IF/00028/2014, 
        and POPH/FSE (EC) by FEDER funding through the ``Programa Operacional de Factores de Competitividade - COMPETE''. This work was supported by Funda\c{c}\~ao para a Ci\^encia e a Tecnologia (FCT) through the research 
        grant UID/DIS/04434/2013 (POCI-01-0145-FEDER-007672) PTDC/FIS-AST/7073/2014 and PTDC/FIS-AST/1526/2014.

\end{acknowledgements}

%-------------------------------------------------------------------

\begin{appendix}
        \section{Nitrogen abundances for stars with and without planets}

        \begin{table*}[!htbp]
                \begin{center}
                        \caption{Nitrogen abundances for a set of stars with planets.}
                    \label{tablacon}
                        \begin{tabular}[ l c c  c c c c]{ l  c  c c c c c c}
                                \hline \hline
                                Star    &       $\rm T_{\rm eff}$ (K)   &       $log g$ (cm $s^{-2}$) & $\xi_{t} (km s^{-1})$        &       [Fe/H]  &       [N/H]         \\                      \hline 
HD 142   &    6431   &    4.82   &    2.1       &    0.05   &    0.12   $\pm$    0.10   \\
HD 1237     &    5489   &    4.46   &    1.04   &    0.06   &    0.07   $\pm$    0.10    \\
HD 2039     &    5990   &    4.56   &    1.24   &    0.34   &    0.36   $\pm$    0.10    \\  
  HD 2638     &    5169   &    4.41   &    0.66   &    0.12   &    0.13   $\pm$   0.11   \\
   HD 4203     &    5728   &    4.23   &    1.18   &    0.43   &    0.53   $\pm$ 0.11   \\
  HD 4208     &    5600   &    4.41   &    0.88   &    -0.29   &    -0.32   $\pm$    0.12 \\
  HD 16141     &    5786   &    4.17   &    1.1   &    0.15   &    0.18  $\pm$   0.10     \\
  HD 17051     &    6237   &    4.46   &    1.31   &    0.2   &    0.29   $\pm$   0.13   \\
  HD 19994     &    6315   &    4.44   &    1.66   &    0.27   &    0.38   $\pm$    0.12  \\
  HD 20794     &    5398   &    4.41   &    0.7   &    -0.41   &    -0.44   $\pm$    0.11 \\
  HD 23079     &    6009   &    4.5   &    1.2   &    -0.11   &    -0.16   $\pm$  0.10 \\
   HD 27894     &    4833   &   4.3   &    0.33  &      0.26    &    0.07   $\pm$    0.13 \\
  HD 28185     &    5621   &    4.36   &    0.92   &    0.19   &    0.24   $\pm$    0.12 \\
  HD 30177     &    5601   &    4.34   &    0.89   &    0.37   &    0.47   $\pm$    0.14  \\
  HD 39091     &    5991   &    4.4   &    1.09   &    0.09   &    0.14   $\pm$   0.10     \\
  HD 50554     &    6129   &    4.41   &    1.11   &    0.01   &    0.10   $\pm$   0.12  \\
  HD 52265     &    6167   &    4.44   &    1.28   &    0.24   &    0.34   $\pm$    0.12  \\
  HD 65216     &    5614   &    4.46   &    0.81   &    -0.17   &    -0.17   $\pm$    0.10 \\
  HD 63454     &    4756   &    4.32   &    0.31  &     0.13    &    0.01   $\pm$    0.11 \\  
  HD 69830     &    5370   &    4.38   &    0.67   &    -0.07   &    -0.06   $\pm$    0.10  \\
  HD 70642     &    5659   &    4.43   &    0.81   &    0.17   &    0.24   $\pm$  0.11    \\
  HD 72659     &    5926   &    4.24   &    1.13   &    -0.02   &    0.08   $\pm$    0.10  \\
  HD 73256     &    5465   &    4.36   &    1.0   &    0.21   &    0.26   $\pm$   0.10   \\
  HD 74156     &    6099   &    4.34   &    1.38   &    0.16   &    0.26   $\pm$  0.11   \\
  HD 75289     &    6139   &    4.35   &    1.22   &    0.3   &    0.26   $\pm$   0.12   \\
  HD 82943     &    5992   &    4.42   &    1.06   &    0.28   &    0.29   $\pm$    0.13   \\
  HD 93083     &    5048   &    4.32   &    0.81   &    0.08   &    0.09   $\pm$    0.10   \\
  HD 106252    &    5880   &    4.4   &    1.13   &    -0.07   &    0.00   $\pm$    -0.10   \\
   HD 114386     &    4774   &  4.37   &    0.01  &     -0.09   &    -0.12   $\pm$    0.10 \\
  HD 114729    &    5865   &    4.2   &    1.29   &    -0.26   &    -0.26   $\pm$    0.10  \\
  HD 117207    &    5649   &    4.31   &    0.95   &    0.21   &    0.35   $\pm$    0.10   \\
  HD 117618    &    6003   &    4.45   &    1.16   &    0.03   &    0.16   $\pm$    0.10   \\
  HD 11964A    &    5285   &    3.81   &    0.95   &    0.06   &    0.06   $\pm$    0.11   \\
  HD 208487    &    6172   &    4.54   &    1.22   &    0.1   &    0.16   $\pm$   0.11   \\
  HD 210277    &    5470   &    4.26   &    0.9   &    0.15   &    0.16   $\pm$   0.10   \\
  HD 213240    &    5967   &    4.28   &    1.22   &    0.13   &    0.15   $\pm$    0.10   \\
  HD 216435    &    6034   &    4.21   &    1.27   &    0.27   &    0.38   $\pm$    0.12   \\
  HD 216437    &    5882   &    4.25   &    1.25   &    0.25   &   0.39   $\pm$   0.11   \\
  HD 216770    &    5351   &    4.31   &    0.85   &    0.2   &    0.29   $\pm$   0.10   \\
  HD 217107    &    5679   &    4.32   &    1.15   &    0.35   &    0.39  $\pm$   0.10   \\
  HD 222582    &    5781   &    4.37   &    1.02   &    -0.01   &    0.05   $\pm$    0.10  \\
          HD 330075     &    4958   &   4.24   &    0.32  &     0.05    &    -0.05   $\pm$    0.10 \\     
\hline
                        \end{tabular}
                        
                \end{center}

        \end{table*}

        \begin{table*}[!htbp]
                \begin{center}
                        \caption{Nitrogen abundances for a set of stars without planets (comparison sample).}
                        \label{tablasin}
                        \begin{tabular}[l c c c  c c c]{ l  c  c c  c c c}
                                \hline
                                Star    &       $\rm T_{\rm eff}$ (K)   &       $log g$ (cm $s^{-2})$ & $\xi_{t} (km s^{-1})$        &       [Fe/H]  &       [N/H]            \\  \hline
                                \hline 
   HD 870    &    5360   &    4.4   &    0.79   &    -0.12   &    -0.14   $\pm$   0.10    \\
 HD 1581     &    5990   &    4.49   &    1.24   &    -0.18   &    -0.23   $\pm$    0.10  \\
  HD 3823     &    6054   &    4.37   &    1.44   &    -0.26   &    -0.36   $\pm$    0.12 \\
    HD 8326     &    4834   &   4.35   &    0.44  &     0.04    &    -0.05   $\pm$    0.10 \\    
    HD 8389A     &    5182   &  4.33   &    0.81   &    0.36   &    0.23   $\pm$    0.11   \\
  HD 9796     &    5139   &    4.34   &    0.49   &    -0.25   &    -0.32   $\pm$    0.10  \\
  HD 15337     &    5088   &    4.36   &    0.51   &    0.06   &    0.01   $\pm$    0.10   \\
    HD 16270     &    4583   &  4.23   &    0.16  &     0.14    &    0.02   $\pm$    0.11 \\
  HD 20807     &    5875   &    4.5   &    1.15   &    -0.23   &    -0.17   $\pm$    0.10 \\
  HD 21019     &    5468   &    3.93   &    1.1   &    -0.45   &    -0.42   $\pm$    0.12 \\
  HD 33636     &    5994   &    4.71   &    1.79   &    -0.08   &    -0.15   $\pm$    0.11 \\
   HD 35854     &    4808   &   4.35   &    0.16  &     -0.14   &    -0.37  $\pm$    0.11 \\
  HD 37226     &    6178   &    4.16   &    1.61   &    -0.12   &    -0.12   $\pm$    0.10  \\
  HD 40105     &    5064   &    3.69   &    0.91   &    0.02   &    0.10   $\pm$    0.10   \\
   HD 44573     &    4990   &   4.42   &    0.61  &     -0.07   &    -0.19  $\pm$    0.11 \\
    HD 65907A    &    5995   &  4.62   &    1.18   &    -0.29   &    -0.24   $\pm$    0.10 \\
  HD 72769     &    5587   &    4.3   &    0.86   &    0.29   &    0.38   $\pm$   0.11    \\
  HD 73121     &    6083   &    4.27   &    1.33   &    0.09   &    0.15   $\pm$    0.11 \\
  HD 76151     &    5781   &    4.44   &    0.93   &    0.12   &    0.21   $\pm$    0.10 \\
  HD 103891    &    6072   &    4.05   &    1.5   &    -0.19   &    -0.08   $\pm$    0.11 \\
  HD 108063    &    6081   &    4.11   &    1.54   &    0.55   &    0.74   $\pm$    0.15 \\
  HD 119629    &    6250   &    4.17   &    1.73   &    -0.17   &    -0.16   $\pm$    0.11 \\
  HD 141597    &    6285   &    4.38   &    1.23   &    -0.4   &    -0.26   $\pm$    0.14  \\
  HD 191033    &    6206   &    4.47   &    1.35   &    -0.19   &    -0.08   $\pm$    0.13 \\
  HD 205536    &    5418   &    4.36   &    0.79   &    -0.07   &    -0.16   $\pm$    0.11 \\
  HD 208068    &    6007   &    4.64   &    1.17   &    -0.38   &    -0.47   $\pm$    0.13 \\
  HD 211415    &    5864   &    4.42   &    1.01   &    -0.21   &    -0.27   $\pm$    0.12 \\
    HD 213042     &    4670   & 4.22   &    0.35  &     0.14    &    0.08   $\pm$    0.11 \\
    HD 214094    &    6288   &  4.28   &    1.46   &    -0.01   &    -0.01   $\pm$    0.10  \\
  HD 220367    &    6116   &    4.45   &    1.44   &    -0.23   &    -0.20   $\pm$    0.11 \\
  HD 222335    &    5220   &    4.48   &    0.62   &    -0.21   &    -0.25   $\pm$    0.10 \\
  CD-436810    &    6011   &    4.41   &    1.09   &    -0.44   &    -0.31   $\pm$    0.11 \\

                                \hline
                        \end{tabular}

                \end{center}
                
        \end{table*}
        
\end{appendix}

\end{document}